# Electron correlation induced small anisotropy in iron-based superconductors

Hyo Seok Ji, Geunsik Lee and Ji Hoon Shim*

*Department of Chemistry, Pohang University of Science and Technology, Pohang 790-784, Korea*

We have investigated the electron correlation effect on the electronic structures and transport properties of the iron-based superconductors using the density functional theory (DFT) and dynamical mean field theory (DMFT). By considering the Fe 3d electron correlation using the DMFT, the quasi-particle bandwidth near the Fermi level is substantially suppressed compared to the conventional DFT calculation. Because of the different renormalization factors of each 3d orbital, the DMFT gives considerably reduced electrical anisotropy compared to the DFT results, which explains the unusually small anisotropic resistivity and superconducting property observed in the iron-based superconductors. We suggest that the electron correlation effect should be considered to explain the anisotropic transport properties of the general d/f valence electron system.

Discovery of high $T_c$ superconductivity (HTSC) in LaFeAsO$_{1-x}$F$_x$ with $T_c$ = 26 K [1] stimulates the effort to understand the mechanism of the HTSC and search for new materials with the HTSC. Similar to the well known cuprate superconductors, it has layered structure with the square lattice of transition metals and the superconductivity emerges from the boundary of the antiferromagnetic ordering of 3d valence electrons. This indicates that the magnetic ordering or more likely its fluctuation is closely related to the electron pairing both in cuprate and iron-based superconductors. Since the first discovery, there have been reported many subsequent iron-based superconductors coming from the parent compounds of ReFeAsOF (Re: rare earth elements), AeFe$_2$As$_2$ (Ae: Ba, Sr, and Ca), LiFeAs, FeSe [2-6] etc. All the superconductors have similar crystal structures and electronic properties. Two dimensional FeAs (or FeSe) layers are commonly appeared and sandwiched by the insulating blocking layers, which induces the anisotropic electronic structures. In all the compounds, the multiple Fe 3d orbitals participate in the magnetic or superconducting transitions. Because of the numerous investigations on the iron-based superconductors, one could use many available experimental data of their physical properties for the comparative study of these materials in clarifying the origin of the HTSC.

The anisotropy is an important parameter to understand the mechanism of the HTSC. Usually the low dimensionality has been believed to be the crucial factor for increasing the superconducting $T_c$. Most of the compounds with the HTSC have layered structure, and the maximum $T_c$ among the cuprate superconductors is designed from the highly anisotropic compounds. Indeed, the density functional theory (DFT) calculations predicted two-dimensional electronic structures and anisotropic transport properties of the iron-based superconductors [7]. With the same strategy to the cuprate superconductors, people tried to synthesize highly anisotropic iron-base superconductor such as Sr$_2$VO$_3$Fe$_2$As$_2$ with a blocking layer of Sr$_2$VO$_3$ [6]. However, the observed maximum $T_c$ in this family is ~37 K [6], which is still below the $T_c$ of less anisotropic compound such as Gd$_{1-x}$Th$_x$FeAsO (56K) [8]. The correlation between the anisotropy and the $T_c$ is not still clarified in the iron-based superconductors. Also recent experimental results report that the observed anisotropy is unusually smaller than the value obtained from the DFT calculations [9]. In order to understand the microscopic mechanism of the superconductivity and its relation to the dimensionality, one needs to predict the anisotropic electronic structures of real materials correctly. In this study, we show that the DFT method systematically overestimates the anisotropy of the iron-based superconductors. Also we suggest that the 3d electron correlation effect should be considered to describe the experimentally observed small anisotropy of the iron-based superconductors correctly.

To investigate the electron correlation effects on the band structures and the electrical anisotropy, we have used the DFT+DMFT approach [10], the combination of the DFT and dynamical mean field theory (DMFT) [11]. The DFT calculations are performed with the full-potential augmented plane-wave method within the WIEN2K code [12], and the DMFT calculations are implemented on top of the DFT Hamiltonian [13]. For the exchange correlation potential in the DFT method, we use the generalized gradient approximation (GGA) in the Perdew-Burke-Ernzerhof [14]. The local electron-electron correlation of Fe 3d orbital is obtained from the DMFT self-consistency equation. The impurity problem within the DMFT equation is solved by the continuous time quantum Monte Carlo (CTQMC) method [15, 16]. In order to show the systematic change of anisotropy without the variation of the interaction parameter, we choose the parameters of Coulomb interaction U=5.0 eV and Hund coupling constant J=0.8 eV in the impurity problem [17]. We perform the calculation with the nonmagnetic phase at the temperature of 72 K. The crystal structure is considered by the experimentally determined lattice constants and internal atomic positions. [1, 4, 5, 18-22]

Using the band structures obtained from the DFT or DFT+DMFT method, one can obtain the electrical contribution of the transport properties. In the Boltzmann equation, the electrical conductivity distribution is calculated as [23, 24]

$$\sigma_{ij} = -\frac{e^2}{4\pi^3}\int \tau v_i v_j \left(\frac{\partial f_0}{\partial E}\right) dk \quad (1)$$

where the group velocity $v_i$ is obtained from the band dispersion $v_i = \partial\varepsilon(k)/\partial k$, the relaxation time $\tau$ is assumed to be constant and $f_0(E)$ is the Fermi-Dirac distribution function [25]. When we consider the anisotropy between x and z directions, the anisotropy of the electrical conductivity is given by Eq. (2) [26].

$$\gamma_\sigma = \sigma_{xx}/\sigma_{zz} \approx v_x v_x / v_z v_z \quad (2)$$

The anisotropy is induced from the anisotropic band structures, which have different band dispersion along each direction. In the s, p orbital system, the DFT method has been successful to describe the transport properties as well as the thermoelectric properties [24].

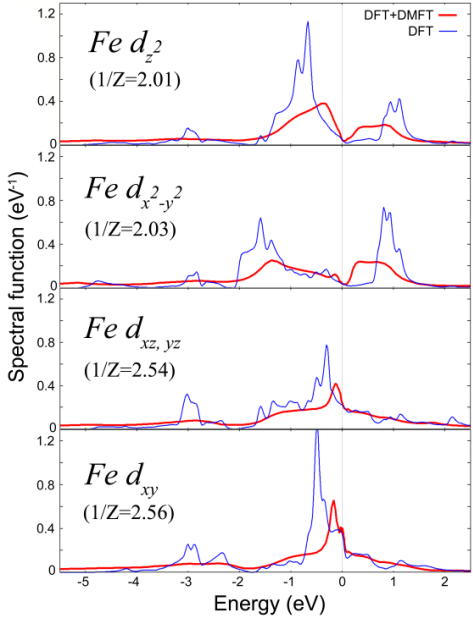

**FIG. 1.** (Color online) Orbital-resolved Fe 3d spectral functions of Ba(Fe$_{0.9}$Co$_{0.1}$)$_2$As$_2$ obtained with the DFT (Blue lines) and DFT+DMFT methods (Red lines).

As an example to represent the electron correlation effects on the spectral function, we show the Fe 3d spectral functions of Ba(Fe$_{0.9}$Co$_{0.1}$)$_2$As$_2$, which has optimal doping for the superconductivity with T$_c$=22 K [19]. We have used the virtual crystal approximation for Co doping on Fe site. All the other iron-based superconductors show similar electronic structures near the Fermi level although there are some variations in the bandwidth and dimensionality due to the different blocking layers. Usually the insulating blocking layers have a role to supply additional electrons or holes to Fe$^{2+}$ valence state and its spectral function has negligible contribution to the Fermi level, so their role in the transport properties is negligible.

In the DFT results as shown in Fig. 1, the spectral functions of d$_{z^2}$ and d$_{x^2-y^2}$ orbitals show well separated peaks around the Fermi level due to the strong covalent bonding between Fe 3d and As 4p orbitals. The main spectral contribution from As 4p orbitals is located between -2 eV and -5 eV. On the other hand, the d$_{xz,yz}$ and d$_{xy}$ orbitals show rather narrow bandwidth and the spectral peak appears near the Fermi level. The contributions of the d$_{xz,yz}$ and d$_{xy}$ orbitals are dominant at the Fermi level, so their electronic structures are important for understanding the transport properties.

In the DFT+DMFT approach, we used the projection-embedding steps between the Kohn-Sham basis and the Wannier-like local basis [13]. In the projection step, the local Green's function is extracted from the full Green's function in Kohn-Sham basis. Then, the local self-energy is obtained from the impurity solver in the Wannier-like local basis. Finally, in the embedding step, the local self-energy is exported back to the Kohn-Sham basis. The DMFT self-consistency condition is given by,

$$G_{loc}(\omega) = \sum_k P_k G(k,\omega) \qquad (3)$$

where P$_k$ is the projection operator which preserves both causality and spectral weight. The full Green's function is given by

$$G(k,\omega) = \frac{1}{\omega + \mu - H_k^{DFT} - E_k \Sigma(\omega)} \qquad (4)$$

where μ is the chemical potential, H$_k^{DFT}$ is non-interacting energy obtained by the DFT calculation, E$_k$ is the embedding operator. The electron correlation effect is treated by the self energy Σ(ω) in the local orbital basis.

Near the Fermi level, quasi-particle bandwidth is substantially suppressed with a factor of 1/Z (=1-∂Σ(ω)/∂ω|$_{ω=0}$) and the incoherent background(Σ(ω)≠0) exists far from the Fermi level as shown in Fig. 1. Usually the bandwidth renormalization factor (1/Z) can be described by the competition of the local Coulomb interaction U and the band width W in one-band Hubbard model. With large W compared to U, the band structures are well described by the conventional DFT method (1/Z~1). As decreasing W/U, the quasi-particle bandwidth is renormalized (1/Z>1) and the spectral weight is transferred to the incoherent background which is called Hubbard bands. Below the critical value of W/U$_c$ (where 1/Z→∞), the spectral weight is totally transferred to the incoherent states and the system becomes the Mott-Hubbard insulator. In the multi-band system such as iron-based superconductors, the Hund's coupling constant J plays an important role in the strength of the electron correlation. The behavior of 1/Z is much more sensitive to J rather than U. With small value of J, there are no well defined Hubbard bands even if U is as big as 5eV. Only with J=0.8eV, 1/Z values of each local 3d orbital becomes 2.0~2.6 in Ba(Fe$_{0.9}$Co$_{0.1}$)$_2$As$_2$. When we slightly increase J, the system shows substantial incoherent spectrum with the suppression of quasi-particle spectrum. The importance of Hund's coupling in the description of metal-insulator transition has been noticed in the two-band Hubbard model at half-filling [27]. In most of the iron-based superconductors, the renormalization factor is estimated by 1/Z=1.5~4 due to the Fe 3d electron correlation [28, 29].

In the anisotropic multi-orbital system, each orbital usually has the different bandwidth W$_i$, so they experience the different renormalization factor 1/Z$_i$ due to the difference of W$_i$/U or W$_i$/J. With smaller bandwidth of d$_{xz,yz}$ and d$_{xy}$ orbital, they have bigger 1/Z$_i$ of 2.54 and 2.56, respectively. With larger bandwidth of the d$_{z^2}$ and d$_{x^2-y^2}$ orbital, they have smaller 1/Z$_i$ of 2.01 and 2.03, respectively. One can also identify the shift of the peak position from the DFT to DFT+DMFT spectral functions, which reflects the different renormalization factors.

The correlation effect can be shown more clearly from the momentum-resolved spectral function. Fig. 2 compares the DFT+DMFT spectral function and the DFT band structures of Ba(Fe$_{0.9}$Co$_{0.1}$)$_2$As$_2$. Due to incoherent part of the spectral function in the DFT+DMFT method, the spectral function cannot be simply interpreted as simple band structures in the DFT method. Near the Fermi level, however, this system follows well the Fermi liquid behaviour, so the spectral function can be clearly defined as a quasi-particle band structures. At the higher energy range, due to the incoherent part of the electron-electron scattering, the spectral functions are blurred and cannot be defined as single band feature. We rescaled the DFT bandwidth by a factor of 2.5, which provides reasonable agreement between the DFT+DMFT spectral functions and the DFT band structures near the Fermi level.

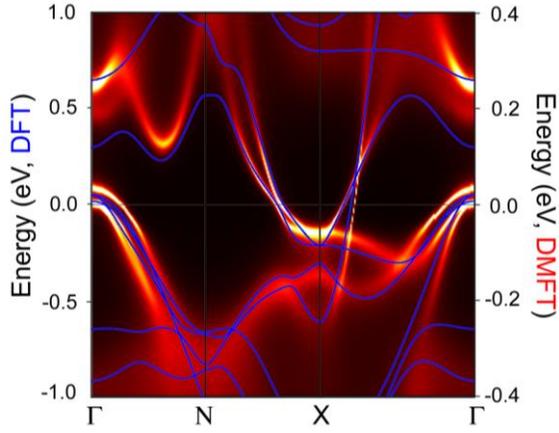

**FIG. 2.** (Color online) The DFT+DMFT spectral function of Ba(Fe$_{0.9}$Co$_{0.1}$)$_2$As$_2$ (Red spectrum) and rescaled DFT band structure (Blue lines) by a factor of 2.5

Due to the different renormalization factors of each Fe 3d orbital, the DFT+DMFT band structures are slightly distorted from the simply rescaled DFT band structures near the Fermi level. The discrepancies are well consistent with the orbital-dependent renormalization factors shown in Fig. 1. Our results are in good agreement with the recently reported ARPES results [30].

Fig. 3 summarizes the electron correlation effect on the conductivity anisotropy schematically. The directional conductivity is calculated by the group velocity along each direction $\sigma_i \sim v_i^2$ as shown in Eq. (1), and its anisotropy is given by $\gamma_\sigma \sim v_x^2/v_z^2$. If we assume that each band has the independent orbital character and they have different renormalization factor $1/Z_i$, they experience the different scaling of the band dispersion and the directional conductivity is given by $\sigma_i' \sim (Z_i v_i)^2$. Finally the conductivity anisotropy has to be modified with the coefficient of $Z_x^2/Z_z^2$. Usually such anisotropic correlation effect is observed in the multi-orbital system such as the d/f valence electron compounds. Indeed, the conventional DFT approaches have failed to explain the experimentally observed anisotropies of the conductivity [9, 31] and superconducting properties in iron-based superconductors, as shown in Fig. 4(a). In the real band structures, the mass enhancement of each band should be understood with the mixture of multi-orbital contribution. Although the 1/Z is a local property for each orbital, the quasi-particle renormalization in the Kohn-Sham basis can have the momentum-dependency $1/Z_k$ by the embedding step. Actually, the Kohn-Sham eigenstates have the multiple characters of the correlated and non-correlated conduction electrons. So, the group velocity, which is the momentum derivative of the eigenvalues, depends on the proportion of each orbitals.

In the DFT calculation, the conductivity anisotropy of Ba(Fe$_{0.9}$Co$_{0.1}$)$_2$As$_2$ is calculated by 7.67, which reflect that this system is rather two dimensional properties, also consistent with the two-dimensional Fermi surfaces in the literature. However, recent experiment on the resistivity reveals that the anisotropy is 2.5~4.5 which is much smaller than the theoretical value obtained from the DFT calculation [9]. Using the DFT+DMFT calculation, the anisotropy is strongly suppressed with the value of 3.24, which is consistent with the experimental value.

We have calculated the anisotropy of superconducting properties of several iron-based superconductors using the DFT and DFT+DMFT method. The anisotropy in the superconducting properties also can be extracted from the band structures according to the Ginzburg-Landau theory. The London penetration depth anisotropy is approximately given by $\gamma_\lambda = \gamma_\sigma^{1/2}$ [26] assuming the isotropic superconducting gap and low temperature limit. In the DFT+DMFT calculation, we use the quasi-particle band structures for the calculation of the group velocity in Eq. (1). The scattering rate coming from the correlation effect is neglected because this system follows well the Fermi liquid behavior at low temperature. If the system is more localized and deviated from the Fermi liquid behavior, the finite size of the scattering rate should be correctly treated in the transport properties.

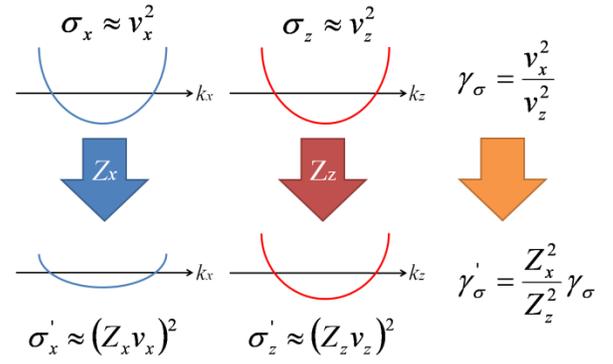

**FIG. 3.** (Color online) Schematic view of the electron correlation effect on the electrical anisotropy($\gamma_\sigma$). $v_i$ is the group velocity obtained from the band dispersion, and $Z_i$ is the band renormalization factor due to the electron correlation effect.

Fig. 4(a) shows the comparison of the penetration depth anisotropies obtained from the DFT calculation and experiments [9, 31-41]. Because various experimental results show different anisotropies for given compounds, we collected all the available experimental results and represent them as the error bar. We assume that the upper critical field anisotropy $\gamma_{H2}(=H_{C2}^{ab}/H_{C2}^{c})$ is same with the $\gamma_\lambda$. Although it shows temperature variance due to the multiple superconducting gaps, we consider them as the error bar. The DFT results clearly show the trends observed in experiments although they systematically overestimate the experimental values. LaFeAsO$_{0.9}$F$_{0.1}$ has the biggest anisotropy due to the biggest blocking layers between FeAs layers. FeSe also has big anisotropy in the calculation due to the empty space. Because its interlayer interaction is very weak due to the van der Waals force, the anisotropy of FeSe is also expected to be big. LiFeAs and BaFe$_2$As$_2$ have the comparable sizes of the anisotropies. With decreasing the size of cation, SrFe$_2$As$_2$ and CaFe$_2$As$_2$ show suppressed anisotropies. For CaFe$_2$As$_2$, the anisotropy is just $\gamma_\lambda$=1.33 with the DFT calculation, which is almost three dimensional. Only CaFe$_2$As$_2$ shows good agreement between the DFT calculation and experimental observations. In all other compounds, the DFT results are above to the experimental values.

By adding the correlation effect with the DFT+DMFT approach, all the iron-based superconductors show substantially reduced anisotropies by factors of 40~80 % which are in good agreement with the experimental results. Although we assume that all the compounds have the same U and J, DFT+DMFT results are within the experimental deviations except SmFeAsO$_{0.75}$F$_{0.25}$. The discrepancy observed in SmFeAsO$_{0.75}$F$_{0.25}$ might be explained by the incorrect use of U and J. More precise experiments on the anisotropy will confirm the importance of the exact estimation of U and J.

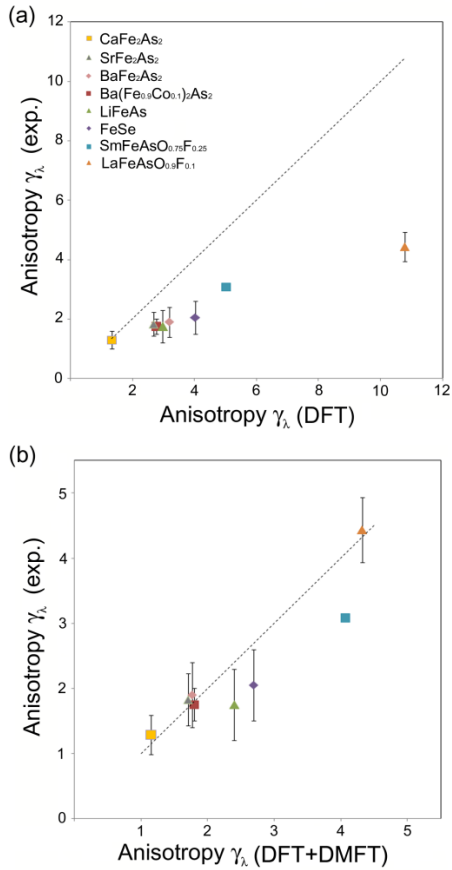

**FIG. 4.** (Color online) The experimental electrical anisotropies [9, 31-41] of iron-based superconductors are compared to the DFT (a) and DFT+DMFT (b) calculation. The dotted lines represent the reference points where the experimental and calculated results are in good agreements.

In the iron-based superconductors, each band has multiple orbital characters, which includes mainly Fe 3d states, so the different $1/Z$ for each 3d orbital gives main contributions in the change of anisotropy. The proportion of As 4p states is also an important factor because they are substantially hybridized with the Fe 3d states near the Fermi level. The change of the anisotropy should be analyzed not only by the different $1/Z$ for each Fe 3d orbital but also the proportion of As 4p orbital. For example, among the two electron Fermi surfaces (FS) near the X point in Fig. 2, one (FS1) has mainly Fe 3d orbital character, which has rather small group velocity due to narrow bandwidth. Another FS (FS2) has a mixed orbital character of Fe 3d and As 4p states, which has bigger group velocity with broad bandwidth. The group velocity of FS1 is more suppressed than that of FS2 because FS1 has more correlated electrons. This difference induces the change in the anisotropy as explained in Fig. 3.

Even though the anisotropy values vary depending on the different lattice constants or blocking layers between FeAs layers [7, 9], all the iron-based superconductors are not so anisotropic as predicted in the DFT calculations. Specially, $AeFe_2As_2$ have almost three dimensional electronic properties [42] as shown in Fig. 4(b). These 3D-like features are clearly different from the quasi-2D copper oxides, so the HTSC in iron-based superconductors should be understood in different way [43]. Therefore we suggest that the correlation effect in the iron-based superconductors should be correctly treated to understand the superconductivity and transport properties of the iron-based superconductors.

In summary, we have used the DFT+DMFT method to show that the electron correlation effect has an important role in describing the anisotropy of the iron-based superconductors. Due to the different renormalization factors of each 3d orbital, the electrical anisotropy by the DFT+DMFT method is substantially suppressed compared to the value of the DFT method. Our results are in good agreement with recent experimental results of the iron-based superconductors.

We acknowledge useful discussions with K. Haule, J. S. Kim and K. H. Kim. Also we appreciate K. Haule for providing DFT+DMFT code [13]. This research was supported by the NRF (No. 2010-0006484, 2010-0026762), WCU through KOSEF (No. R32-2008-000-10180-0), and by the POSTECH Basic Science Research Institute Grant.